\documentclass{article}

\usepackage{PRIMEarxiv}

\usepackage[utf8]{inputenc} 
\usepackage[T1]{fontenc}    
\usepackage{hyperref}       
\usepackage{url}            
\usepackage{booktabs}       
\usepackage{amsfonts}       
\usepackage{nicefrac}       
\usepackage{microtype}      
\usepackage{lipsum}
\usepackage{fancyhdr}       
\usepackage{graphicx}       
\usepackage{subcaption}
\graphicspath{{media/}}     

\title{Regression-based approach to anxiety estimation of spider phobics during behavioural avoidance tasks}

\author{
    Florian Grensing \\
    Data Engineering\\
  Helmut-Schmidt-University \\
  Hamburg\\
  \texttt{Grensinf@hsu-hh.de} \\
   \And
    Vanessa Schmücker \\
    Clinical Psychology\\
    University of Siegen\\
    Siegen\\
    \texttt{Vanessa.Schmuecker@uni-siegen.de}
    \And
    Anne Sophie Hildebrand\\
    Clinic for Psychiatry and Psychotherapy \\
    Bielefeld University\\
    Bielefeld\\
    \texttt{Anne.Hildebrand@evkb.de}
    \And
    Tim Klucken \\
    Clinical Psychology\\
    University of Siegen\\
    Siegen\\
    \texttt{Tim.Klucken@psychologie.uni}
   \And
  Maria Maleshkova \\
  Data Engineering\\
  Helmut-Schmidt-University \\
  Hamburg\\
  \texttt{Maleshkm@hsu-hh.de} \\
}

\begin{document}
\maketitle

\begin{abstract}

Phobias significantly impact the quality of life of affected persons. Two methods of assessing anxiety responses are questionnaires and behavioural avoidance tests (BAT). While these can be used in a clinical environment they only record momentary insights into anxiety measures.
In this study, we estimate the intensity of anxiety during these BATs, using physiological data collected from unobtrusive, wrist-worn sensors. Twenty-five participants performed four different BATs in a single session, while periodically being asked how anxious they currently are.
Using heart rate, heart rate variability, electrodermal activity, and skin temperature, we trained regression models to predict anxiety ratings from three types of input data: (1) using only physiological signals, (2) adding computed features (e.g., min, max, range, variability), and (3) computed features combined with contextual task information.
Adding contextual information increased the effectiveness of the model, leading to a root mean squared error (RMSE) of 0.197 and a mean absolute error (MAE) of 0.041. Overall, this study shows, that data obtained from wearables can continuously provide meaningful estimations of anxiety, which can assist in therapy planning and enable more personalised treatment.

\end{abstract}


\section{Introduction}

Phobias are intense and irrational fears of specific objects, situations, or activities that significantly impact an individual’s daily life \cite{nimh_phobia_examples}. Common phobias include arachnophobia (fear of spiders), acrophobia (fear of heights), and claustrophobia (fear of confined spaces)\cite{nimh_phobia_examples}. 
%
A central component in the treatment of phobias is exposure therapy \cite{JoschaBohnlein.2020}, in which individuals are gradually exposed to the feared stimulus in order to reduce their anxiety over time. A commonly used tool to assess phobia severity in clinical and research settings is the Behavioural Avoidance Test (BAT) \cite{BAT_Definition}. In this task, individuals are asked to approach a fear-inducing stimulus, such as a spider, as closely as possible, but are allowed to stop whenever they wish. The point at which they stop is recorded as a behavioural measure of anxiety intensity. Variants of the BAT include walking toward the stimulus \cite{Mühlberger}, pulling it towards oneself \cite{Schwarzmeier.2020}, or interacting with it in virtual reality (VR) environments \cite{Mühlberger}, which allow for more controllable and replicable exposure scenarios. Although the BAT is not typically part of exposure therapy itself, it is frequently used to evaluate treatment outcomes before and after therapeutic interventions \cite{Castagna.2017}.

While the BAT provides valuable insights into anxiety behaviour, it does not offer insight into the fluctuations of anxiety during the task or throughout treatment. Recent advancements in physiological monitoring offer new methods for objectively assessing fear and anxiety. Measures such as heart rate and skin conductance have been shown to correlate with emotional arousal, and could provide a more continuous, real-time estimate of anxiety during exposure sessions \cite{WataruSato.2020}\cite{SudieE.Back.2022}.

In this study, we investigate how physiological signals can be used to estimate moment-to-moment anxiety intensity of subjects with a fear of spiders. Specifically, we aim to model self-reported anxiety during BATs using data such as heart rate and skin conductance. Our approach offers a step toward more objective, data-driven methods for monitoring patient responses during exposure therapy. To this end, we collect and analyse physiological and self-report data during BATs and develop machine learning models that estimate continuous anxiety intensity.

The contributions of this study are (1) the collection of data from 25 participants during four Behavioural Avoidance Tests, (2) estimation of continuous anxiety intensity using wrist-worn wearables, and (3) integration of contextual information into the machine learning models. 

This paper is structured as follows: First, we present some background on related works. Afterwards, we describe the study setup, as well as the data collection and preprocessing. Next, the results will be presented and discussed. Finally, the results are summarised and an outlook for future research is presented.

\section{Research Background}\label{researchBackground}


Fear responses are mediated by the autonomic nervous system (ANS), particularly the sympathetic division, which triggers the “fight-or-flight” response \cite{Kreibig.2010}. This then triggers various physiological responses, which can be used for emotion detection. 
%
Although works, such as that of Ekman et al. \cite{Ekman} suggest using camera-based facial recognition, this is often difficult in combination with technologies like virtual reality, which are becoming increasingly popular in the field of therapy. The same is true for electroencephalography (EEG), which, while informative, is intrusive and can be difficult to integrate into VR-based applications. Therefore, this paper specifically focuses on physiological signals that are measurable by using wrist-worn wearables, such as heart rate (HR), electrodermal activity (EDA), also known as galvanic skin response (GSR) and body temperature.

These signals have successfully been used for emotion detection in the past. EDA has been found suitable for emotion classification among both the valence and arousal dimensions of emotion recognition \cite{EDAVR2}. A study by Polo et al. has used EDA to distinguish between sadness, happiness, fear and relaxation \cite{EDAVR2}. Ménard et al. have used a combination of EDA and heart rate, to classify emotions \cite{emotionHR}. Participants were shown video clips to elicit different emotions, which were then classified with an accuracy of 85\% for EDA and 89\% for heart rate respectively. Other signals, such as heart rate variability, breathing rate, blood pressure have also been used \cite{EGGER201935}.


One study that used physiological signals to estimate anxiety during virtual reality therapy was performed by Freeman et al. \cite{DanielFreeman.2018} and used VR for treating fear of heights. During the experiment, EEG, GSR and respiration data were collected. Sixty-one participants performed a relaxation–stress–relaxation type tasks multiple times over four weeks. Using the collected physiological data, the authors achieved an 88\% accuracy for a binary classification using a support vector machine (SVM).

In 2021, Šalkevicius et al. presented their work, in which they collected BVP, GSR and skin temperature data during a VRET and used that data to predict the participants anxiety levels. They split the anxiety into 4 classes (‘low’, ‘mild’, ‘moderate’, and ‘high’) and performed classification. Across 30 participants, they have attained an accuracy of 80.1\% \cite{Salkevicius.2019}.

While these studies have demonstrated that physiological signals can be used to classify emotional states, relatively few have focused on predicting continuous, moment-to-moment anxiety intensity using wearable data. Šalkevicius et al. have incorporated Subjective Units of Distress Scale (SUDS) ratings in the context of virtual reality exposure therapy (VRET), they have discretized these scores into categorical anxiety levels (e.g., low, mild, moderate, high). In contrast, the present study uses regression to predict continuous SUDS scores, allowing for a more fine-grained analysis of anxiety fluctuations over time.

\section{Methods}

In the following section, we describe the experiment design, including participant recruitment, the four Behavioural Avoidance Tests, the preprocessing of the physiological data and the machine learning models used for the anxiety prediction.

\textbf{Participants:} The study involved 25 participants (3 male, 22 female) with an average age of 24.96 years (SD = 8.75). All participants met the inclusion criteria of being between 18 and 65 years old and having a specific phobia of animals, specifically spiders. Exclusion criteria included a lifetime diagnosis of substance use disorder, bipolar disorder, or psychotic disorder, as well as current severe depressive episodes, suicidality, use of medications such as benzodiazepines or barbiturates, or neurological conditions such as epilepsy. Full inclusion and exclusion criteria are detailed in our earlier concept paper \cite{10896228}.

Unfortunately, the data of two participants had to be discarded due to technical problems with the sensor, leaving 23 participants for the evaluation.

\vspace{0.5em}

\textbf{Ethics Statement} This study was funded by the Research Training Group 2493 [2493/1: 398510439],
German Research Foundation (DFG). The funder did not have a role in the design, data collection, data
analysis, and reporting of this study. Authors state no conflict of interest. Informed consent has been obtained from all individuals included in this study. The research related to human use complies with all the relevant national regulations, institutional policies and was performed in accordance with the tenets of the Helsinki Declaration, and has been approved by the ethics commission of the University of Siegen, Germany, reference number: ER\_04\_2023.
\vspace{0.5em}

\textbf{The BATs}
For this study, we have developed four BATs: \textbf{Walk-BAT}, \textbf{VR-Walk-BAT}, \textbf{Table-BAT} and \textbf{VR-Table-BAT}. These are described in more detail in our concept paper \cite{ourSpider}. Each of these BATs was designed to elicit anxiety from participants with spider phobia by using either a real or a virtual spider.

In the \textbf{Walk-BAT}, participants enter a room and walk towards a spider on the other side of the room. For the \textbf{Table-BAT} on the other hand, participants sit down in front of a table with a crank-device. The spider is then carried into the room and placed on the far end, on a movable platform. The participants then use the crank device to pull the spider towards themselves. 

\textbf{VR-Walk-BAT} and \textbf{VR-Table-BAT} are virtual reality adaptations designed to be as similar to their in vivo counterparts as possible. 

During each of these BATs participants give multiple subjective anxiety ratings, either verbally in vivo, or via a controller input in virtuo. Anxiety ratings are given between 0 (no anxiety) and 100 (extreme anxiety). 
These anxiety ratings are collected: 
\begin{itemize}
    \item before the task explanation
    \item before a BAT (anticipation)
    \item at the beginning of the BAT
    \item for every 25\% of distance to the spider covered
    \item when the final distance is reached
    \item after the participant or the spider has left the room
\end{itemize}

Each participant performs all four BATs in a single session. Between the BATs, the participants fill out various questionnaires. This serves to collect psychometric data, but also to give the participants some time to calm down and their physiological signals to return to normal.  
Physiological signals are collected throughout the entire session using an Empatica E4 \cite{Empatica} wristband. 
\vspace{0.5em}

\textbf{Hard- and Software}
The BATs in virtuo were designed using the Unity Game Engine \cite{Unity} (version 2021.3.2f1) and optimized for the HTC Vive Pro \cite{VivePro}. This setup ensures that once the room measurements are completed, the size and orientation of the VR-Area are consistent for all participants. 
To remain compatible with this and other possible VR setups, we have decided against using facial recognition and EEG, as these setups require significant overhead and may conflict with VR setups. For this reason, the wrist-worn Empatica E4 sensor \cite{Empatica} was chosen. This wrist-worn sensor collects EDA, HR, IBI as well as skin temperature, giving it a wide range of sensors, appropriate for emotion recognition.

In order to synchronise the anxiety ratings with sensor data, we used a custom-built synchronisation framework \cite{FRAME}. This tool allowed experiment supervisors to input the anxiety ratings, as well as when which BAT is started and finished. These events are then time-stamped and digitised, allowing the physiological data to be synchronized with the observation data. The data is then loaded into python for preprocessing and post-hoc analysis.
\vspace{0.5em}

\textbf{Preprocessing}
 The preprocessing of EDA data followed best practice recommendations as described by Benedek et al. \cite{EDApreprocessing} and others. First, we interpolated any individual missing values. Then we applied a low-pass Butterworth filter with a cutoff of 0.5Hz. Afterwards, the EDA signal was decomposed using the NeuroKit2 \cite{NeuroKit} python package using a high-pass filter. This provides us with the slow-changing tonic EDA as well as the fast-changing phasic EDA.

To extract HRV features from wearable IBI data, we applied a 5-minute sliding window with a 0.25-second step size. Implausible IBI values (below 300 ms or above 2000 ms) were removed to reduce the impact of artefacts. Following this, we computed two standard time-domain HRV features: the standard deviation of all IBI values in the window (SDNN), which reflects the overall heart rate variability, and the root mean square of successive differences (RMSSD), which primarily shows short-term activity.

While the distance to the spider were available continuously during the BATs in virtuo, for the ones in vivo, the distances were only available at the start, end, and every 25\% distance covered. As such, the distance data in vivo, as well as the anxiety ratings across all four BATs, are linearly interpolated. Similarly, the heart rate and skin temperature data were also upsampled from 1Hz to 4Hz, to align better with the higher sampling rate of the EDA sensor.

This leaves us with the following features, which can be used for analysis:

\begin{itemize}
    \item Target: Anxiety
    \item Heart rate (HR)
    \item HRV\_SDNN
    \item HRV\_RMSSD
    \item Skin temperature
    \item EDA tonic
    \item EDA phasic
    \item Current step of the experiment (which BAT or questionnaire)
\end{itemize}

\textbf{Feature Selection and Engineering}
 All values were normalized per participant using a min-max scaler to a [0,1] range. The same scaling was applied to the anxiety values, so that the highest anxiety value of each participant was scaled to 1, and the minimum to 0.

Then we calculated the correlation between the potential inputs and the self-reported anxiety. In addition to the current physiological values, the difference and ratio to the baseline (measured at the start of the session) were used. Finally, statistics computed over a sliding window of 1, 5, 10 and 30 seconds were calculated. These include minimum, maximum, range, mean, standard deviation and change.

\begin{figure}[h] 
    \centering
    \includegraphics[width=1.2\textwidth]{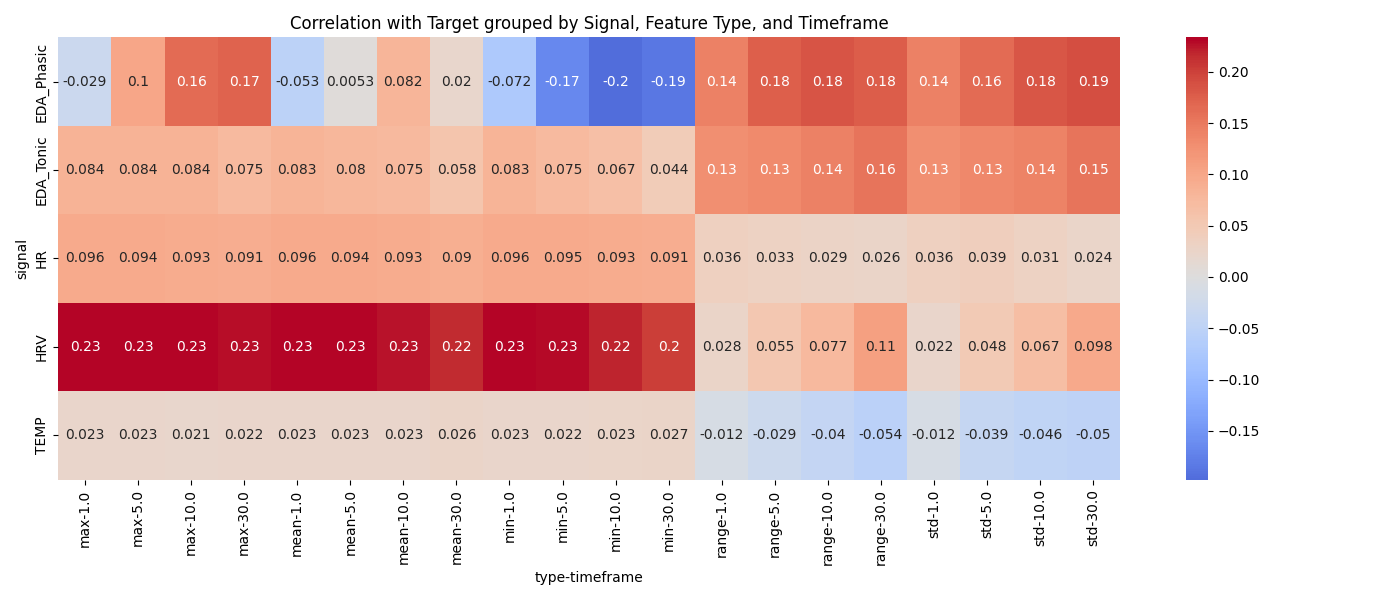} 
    \caption{Correlations of statistical measures}
    \label{fig:correlation1}
\end{figure}

These correlations for the statistical measures can be seen in figure~\ref{fig:correlation1}. To avoid redundancy, only one time window should be used. The tendency here clearly shows that the correlation is higher for the 30-second windows. Additionally, only a very low correlation is seen for all temperature values. Due to this, we decided to only use the 30-second window for computed features.  

In addition to physiological signals, we also considered using contextual information. Here, the current stage of the experiment was encoded into three binary variables: \textbf{isBAT}, indicating whether any BAT is currently being performed; \textbf{isWalk}, indicating whether the current BAT involves walking; and \textbf{isVR}, indicating whether the current BAT is performed in VR. 

The highest correlation among these was found for isBAT (0.64) likely due to the very low anxiety levels when no BAT is taking place. Furthermore, isWalk (0.37) and isVR (0.25) are relevant for investigating, how the body responds to the different BAT types.

\textbf{Used Regression Models}

For this experiment, we deployed three common machine learning algorithms for time-series data. As a baseline, we used a random forest regression algorithm with n = 50.  
In addition to this, XGBoost with n\_estimators=100 was used. 
%
Lastly, we also used a long short-term memory (LSTM) neural network with the layers LSTM (64), LSTM (32), Dense (16) and a Dense (1) output layer.
As loss functions we used both "MSE" and "Huber". The "Huber" loss function is more reactive to sudden changes, whereas "MSE" provides smoother results. For our experiment, "MSE" has yielded better results, which is why "Huber" will not be described further.
A batch size of 32 and a training time of 50 epochs were used to balance training times and performance.

\section{Results}

For this study, we used regression algorithms to estimate anxiety values. As such, the results are measured using the mean absolute error (MAE) and the root mean squared error (RMSE). 
%
For each of the three machine learning algorithms, we performed a five-fold cross validation using the same splits. 
Each cross validation was repeated three times. The first used only the preprocessed, but not computed features, and will be referred to as \textbf{Basic}. The second (\textbf{Features}) used these signals, but included the computed features. Lastly, contextual data, in the form of the current phase of the experiment, was included (\textbf{Context}). The results of the experiments can be seen in table~\ref{tab:results}.

\begin{figure}[htbp]
    \centering
    \begin{subfigure}[b]{0.45\textwidth}
        \includegraphics[width=1.11\textwidth]{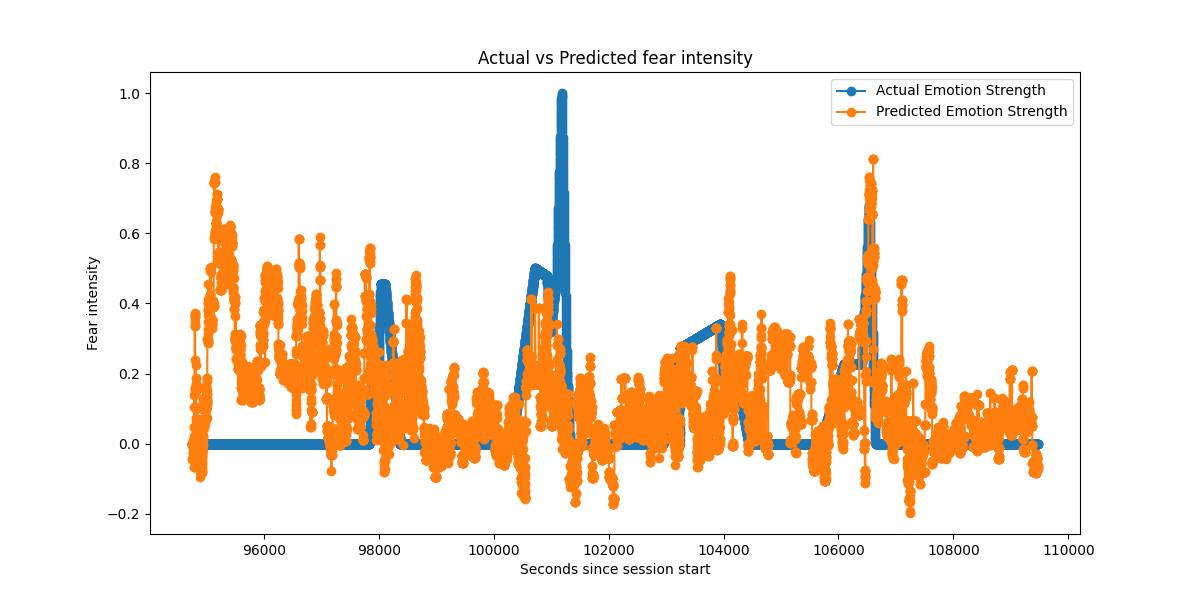}
        \caption{XGBoost Basic}
    \end{subfigure}
    \hfill
    \begin{subfigure}[b]{0.45\textwidth}
        \includegraphics[width=1.11\textwidth]{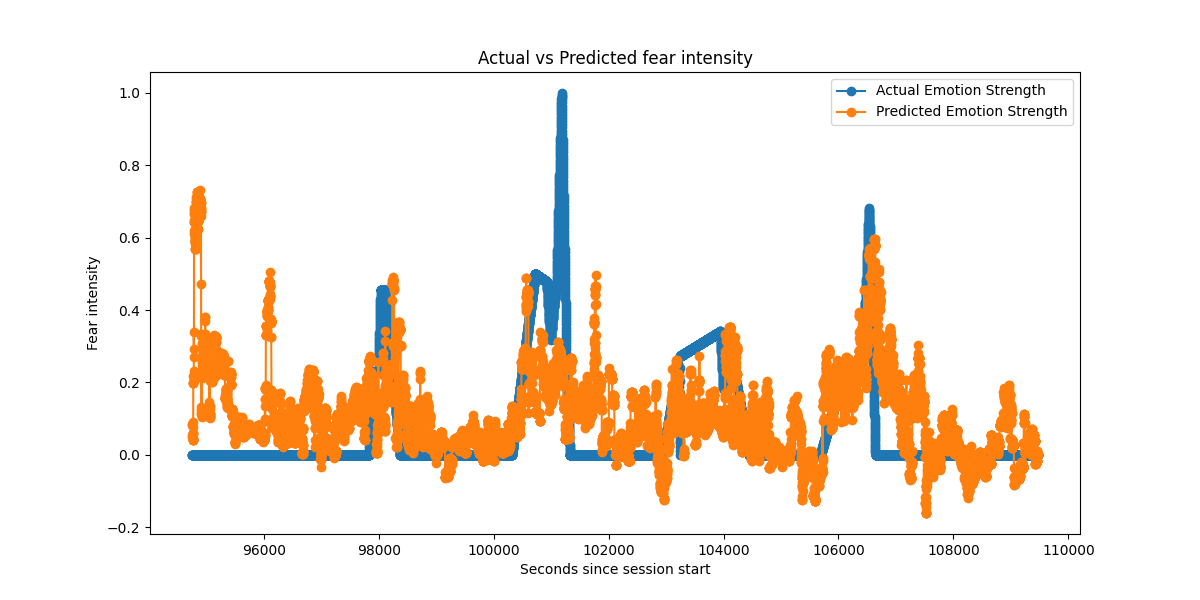}
        \caption{XGBoost Features}
    \end{subfigure}
    \hfill
    \begin{subfigure}[b]{0.45\textwidth}
        \includegraphics[width=1.11\textwidth]{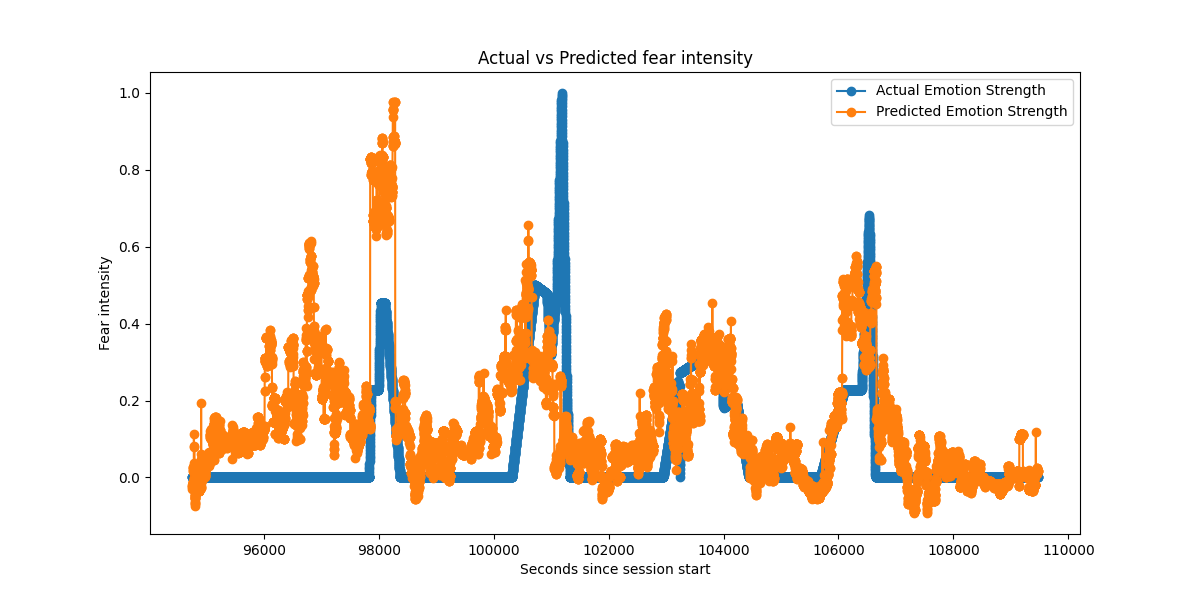}
        \caption{XGBoost Context}
    \end{subfigure}
    \caption{Comparison of Models using XGBoost}
    \label{fig:XGB}
\end{figure}

\begin{figure}[htbp]
    \centering
    \begin{subfigure}[b]{0.45\textwidth}
        \includegraphics[width=1.11\textwidth]{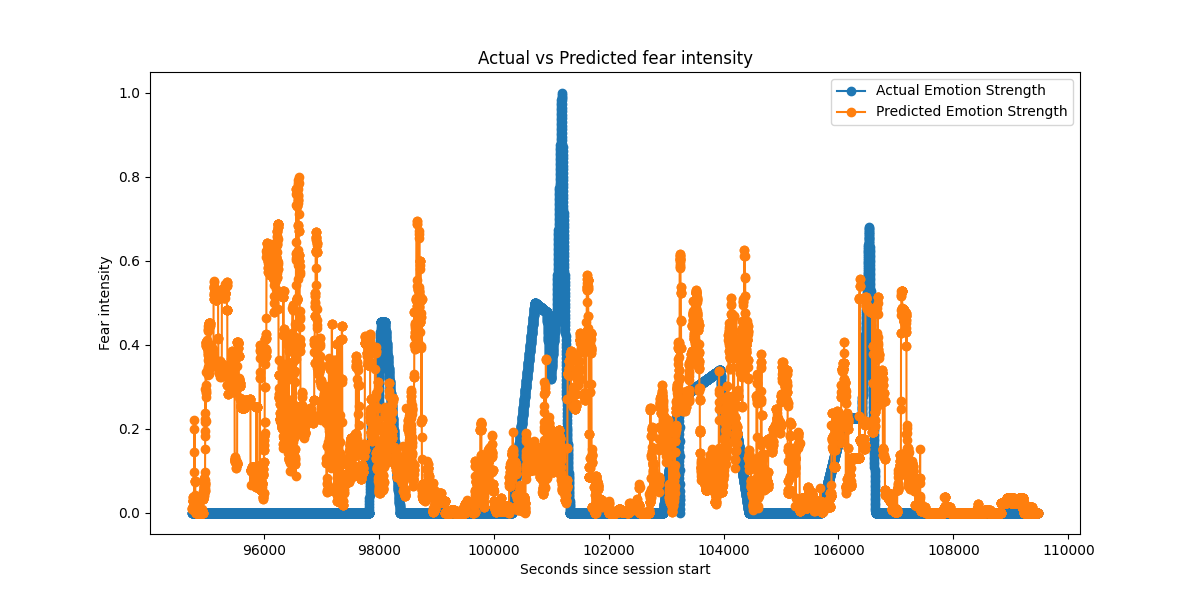}
        \caption{Random Forest Basic}
    \end{subfigure}
    \hfill
    \begin{subfigure}[b]{0.45\textwidth}
        \includegraphics[width=1.11\textwidth]{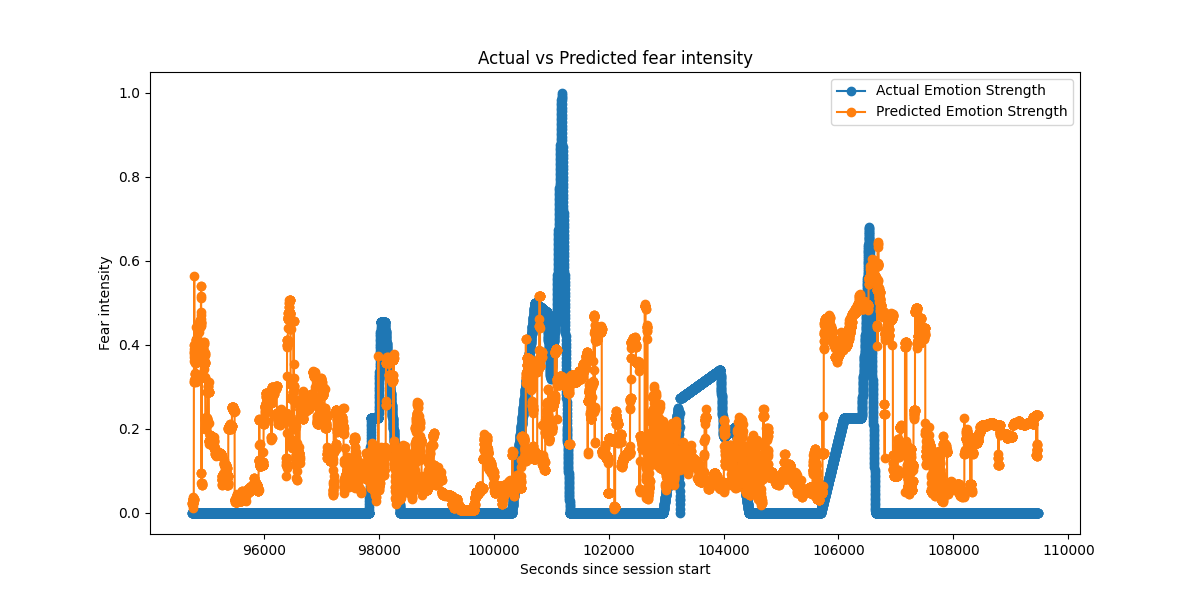}
        \caption{Random Forest Features}
    \end{subfigure}
    \hfill
    \begin{subfigure}[b]{0.45\textwidth}
        \includegraphics[width=1.11\textwidth]{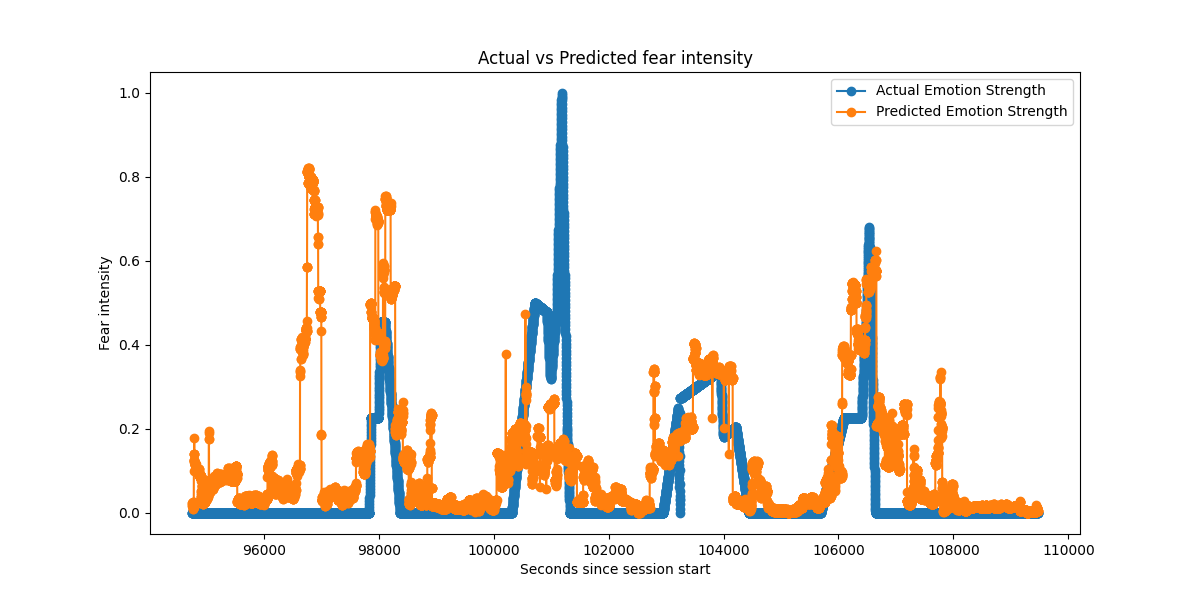}
        \caption{Random Forest Context}
    \end{subfigure}
    \caption{Comparison of Models using Random Forest}
    \label{fig:RFs}
\end{figure}

\begin{figure}[htbp]
    \centering
    \begin{subfigure}[b]{0.45\textwidth}
        \includegraphics[width=1.11\textwidth]{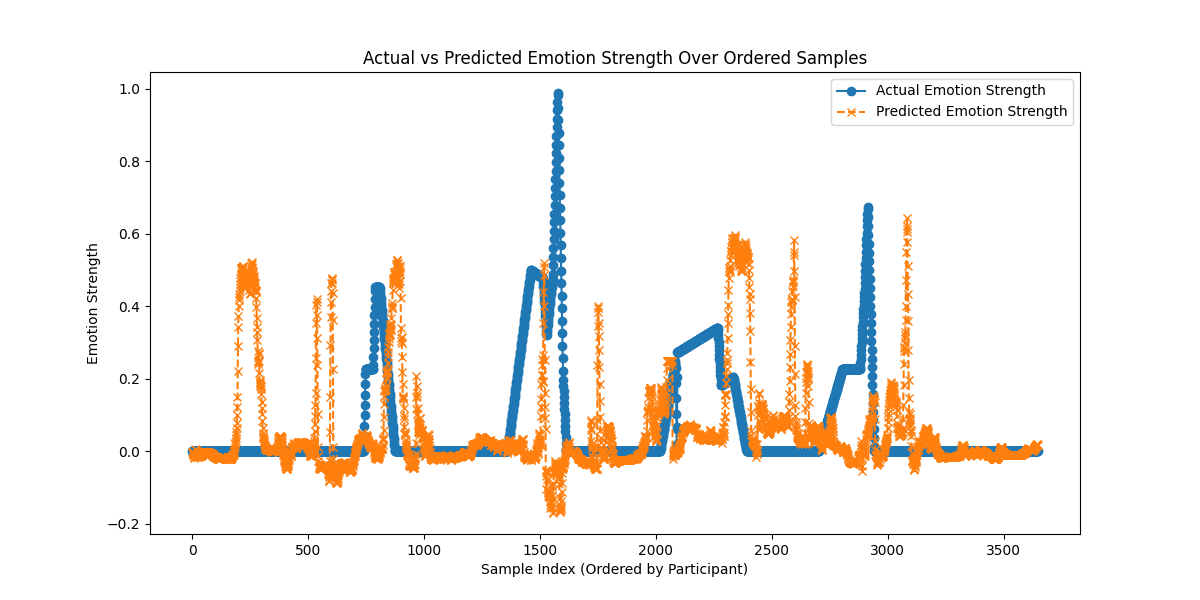}
        \caption{LSTM Basic}
    \end{subfigure}
    \hfill
    \begin{subfigure}[b]{0.45\textwidth}
        \includegraphics[width=1.11\textwidth]{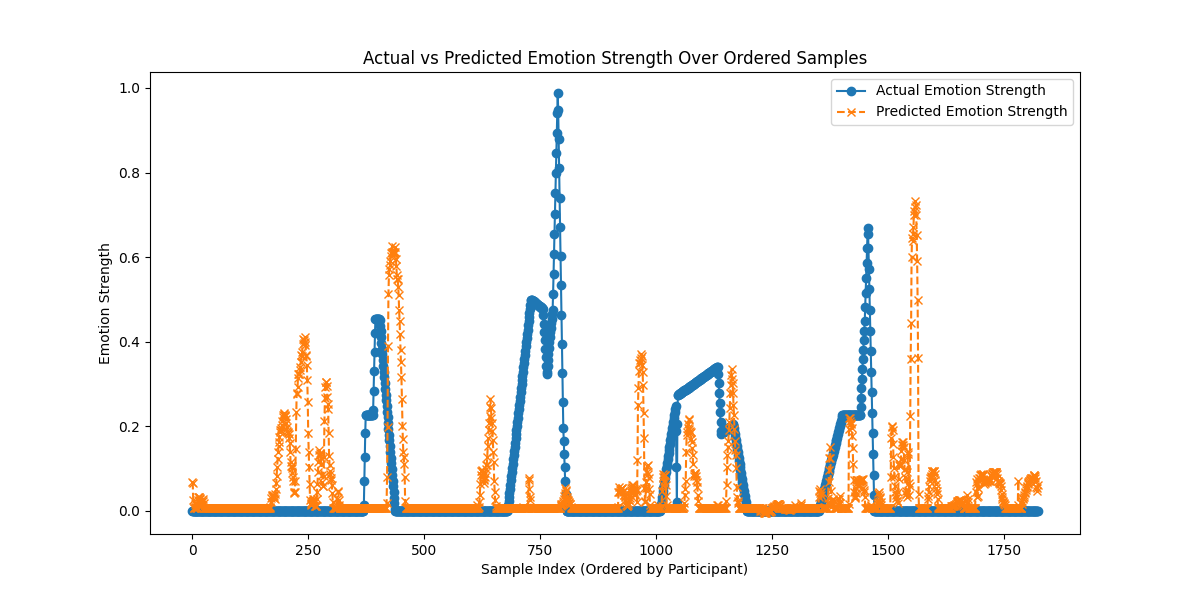}
        \caption{LSTM Features}
    \end{subfigure}
    \hfill
    \begin{subfigure}[b]{0.45\textwidth}
        \includegraphics[width=1.11\textwidth]{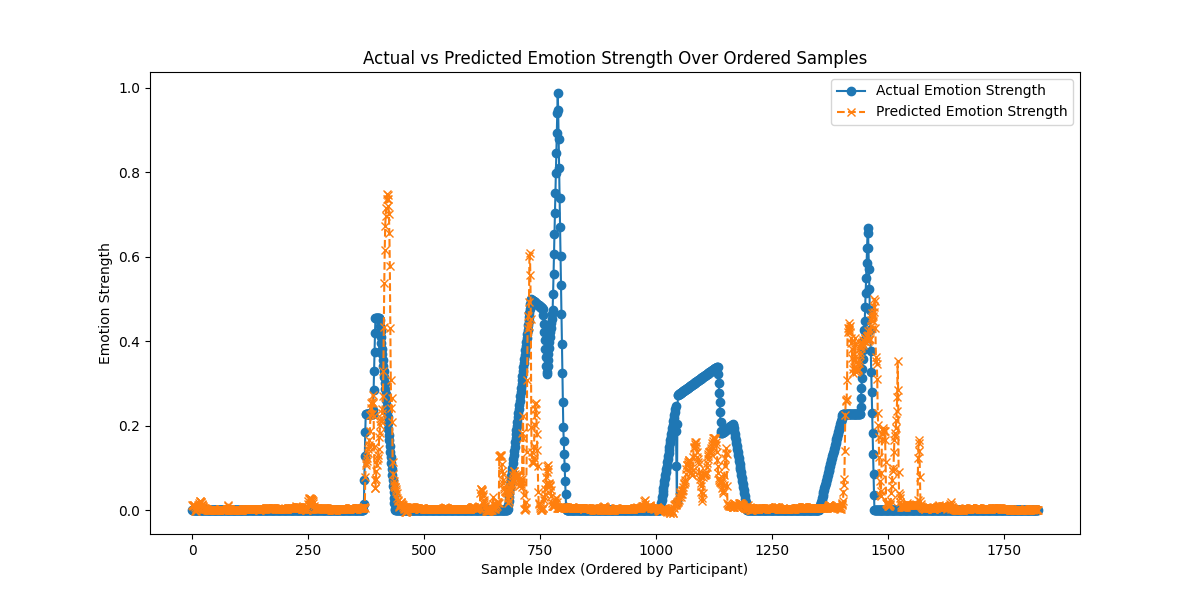}
        \caption{v Context}
    \end{subfigure}
    \caption{Comparison of Models using LSTM}
    \label{fig:LSTMs}
\end{figure}

Figure~\ref{fig:XGB} each show one example result for each of the XGBoost results. Figures~\ref{fig:RFs} and ~\ref{fig:LSTMs} respectively show the same participant for the different models of Random Forests and LSTMs. These figures each show the same participant over the duration of the entire experiment, including all four BATs, as well as questionnaire phases in between. 

An increase in model performance was observed with an increase in input complexity. Using only the preprocessed physiological data resulted in a baseline performance of MAE = 0.078 and RMSE = 0.275. With this, the roughest trends of the anxiety values are approximated; however, there is a lot of noise, especially outside the BATs, as seen in figure~\ref{fig:XGB} (a).

Adding computed features such as the mean and standard deviation of each signal led to a noticeable improvement (MAE = 0.064 and RMSE = 0.249). This leads to significantly less noise and a smoother approximation of the anxiety values, which can be seen in figure~\ref{fig:XGB} (b). 
Finally, by adding context information on the current phase and type of BAT, the best result of MAE = 0.041 and RMSE = 0.197 was achieved. This can be seen in figure~\ref{fig:XGB} (c), and shows that the trends during the BATs are followed very closely, with relatively minor noise, especially during the relaxation periods. 
 
As the anxiety values are zero in between the different tests, they potentially skew the results. As such, we have included a filtered version of the results for both MAE and RMSE, which includes only values during the BATs, not during the resting phases. 

\begin{table}[h]
    \centering
    \caption{Results}
    \label{tab:results}
    \begin{tabular}{lllll}
        \hline
          Method  & MAE & RMSE & MAE (Filtered) & RMSE (Filtered)  \\ 
        \hline
        Random Forest Basic & 0.086 & 0.289 & 0.119 & 0.336\\ 
        Random Forest Features & 0.066 & 0.254 & 9.107 & 0.321 \\ 
        Random Forest Context & 0.045 & 0.209 & 0.085 & 0.28 \\ 
        XGBoost Basic & 0.078 & 0.275 & 0.117 & 0.333\\ 
        XGBoost Features & 0.064 & 0.249 & 0.111 & 0.323 \\ 
        XGBoost Context & 0.041 & 0.197 & 0.077 & 0.269 \\ 
        LSTM Basic & 0.0855 & 0.281 & 0.122 & 0.339 \\ 
        LSTM Features & 0.0869 & 0.289 & 0.143 & 0.367\\ 
        LSTM Context & 0.046 & 0.209 & 0.0956 & 0.295\\ 
        \hline
        
    \end{tabular}
\end{table}

\section{Discussion}

The results show that continuous anxiety can be determined on a moment-to-moment basis using physiological signals. While most studies in this field focus on classifying anxiety into discrete labels (low, medium, high), using regression preserves the full resolution of the anxiety values. 
This allows us to capture more subtle changes, especially those, that would otherwise remain in the same class, while not overstating small changes on the border between two classes. As such, it is a more natural representation of the psychological state of participants.

Furthermore, the smoother output is more suitable more suitable for real-time adaptive systems. They would allow more nuanced responses, such as adapting the intensity of exposition, or offering supportive feedback. In this way, a biofeedback driven therapy would be possible. Physiological signals could be analysed to determine the intensity of anxiety, which in turn can be used directly influence the therapy session, especially in combination with VR-technology.

As most studies focus on classification of emotions, it is challenging to compare our results to those of other works. Even between different works using regression, a comparison is difficult, due to differences in sensors, participants and experiment design. However, our results of RMSE = 0.197 and MAE = 0.041 is comparable to works such as that of Galvão et al., who have achieved an RMSE of 0.16 and MAE of 0.06 for valence and arousal prediction using EEG signals \cite{Galvao.2021}. This further supports the feasibility of regression-based algorithms for anxiety prediction for therapeutic environments.

Additionally, the results also highlight the importance including contextual information for emotion recognition tasks. Not only does the context improve performance by helping the model interpret the data, the contextual information is also typically available in a therapeutic environment. 
This can be simple information, such as if the participant is standing or walking, or information describing the task, such as using VR or the current phase of exposure.

The results might be further improved by addressing a few key challenges. The first is the size of the dataset. Only having data from 23 participants, most of whom were young and female, the results are not representative of the general populace. Another challenge is the individuality of physiological signals. Every participant reacts to the stimulus a little differently. Additionally, the physiological reactions can also be influenced by external factors, such as temperature. While we aimed to keep external influences to a minimum, they can never be fully prevented. This could also be countered by a sufficiently large dataset, as well as robust calibration and normalisation techniques.

A final challenge lies in the subjective anxiety ratings used as ground truth for the experienced anxiety. While all participants used the same 0-100 scale, a value of 50 does not represent the same intensity of anxiety for every participant. This too can be reduced by a large enough dataset, however, it might also be possible to find other data to serve as labels for the training. One possibility would be the current or final distance to the spider. For example, it might be possible, to treat the final distance to the spider as 100, representing an extreme anxiety, and interpolate the distance from the starting point to that final distance as the anxiety value. However, this would not include the anxiety experienced from the anticipation at the beginning of each trial.


\section{Conclusion}

In this study we investigate the feasibility of using physiological signals to continuously determine the intensity of anxiety experienced by participants with a fear of spiders. To do this, the participants performed a row of behavioural avoidance tasks, regularly giving a subjective anxiety rating. These anxiety ratings were then used to train various machine learning models to estimate the anxiety intensity, based on physiological data. The best results were an RMSE of 0.197 (filtered 0.269) and MAE of 0.041 (0.077).

This result was accomplished by combining the physiological data with contextual data about the experiment being performed. This highlights the importance of the context when using such systems, as context will lead to better insight into patient responses, improving care options for therapists. 

In the future, based on our work, we aim to create a virtual reality exposure therapy, which can automatically adjust to the patient response and anxiety levels. This will allow more personalised treatment, improving efficiency, while also reducing the strain placed on the therapist.


\bibliographystyle{plain}
\bibliography{references}

\end{document}